\documentclass[sigconf, 11pt]{acmart}

\usepackage{booktabs} 

\usepackage{macros}

\usepackage[utf8]{inputenc} 
\usepackage[T1]{fontenc}    
\usepackage{hyperref}       
\usepackage{url}            
\usepackage{booktabs}       
\usepackage{amsfonts}       
\usepackage{amsmath}       	
\usepackage{nicefrac}       
\usepackage{microtype}      
\usepackage[capitalise]{cleveref}
\usepackage{graphicx}
\usepackage{breakurl}
\usepackage{soul}
\usepackage{xcolor}

\usepackage{mathtools}
\mathtoolsset{showonlyrefs}

\usepackage[font={footnotesize}]{caption}

\setcopyright{none}

\acmISBN{123-4567-24-567/08/06}
\acmConference[FAT/ML]{Workshop on Fairness, Accountability, and Transparency in Machine Learning}{2017}{Halifax, Nova Scotia} 
\acmYear{2017}
\copyrightyear{2017}

\begin{document}
\title{New Fairness Metrics for Recommendation that Embrace Differences}

\author{Sirui Yao}
\affiliation{%
  \institution{Department of Computer Science}
  \city{Blacksburg} 
  \state{Virginia} 
  \postcode{24060}
}
\email{ysirui@vt.edu}
\author{Bert Huang}
\affiliation{%
  \institution{Department of Computer Science}
  \city{Blacksburg} 
  \state{Virginia} 
  \postcode{24060}
}
\email{bhuang@vt.edu}
%
%
%
%
%
%
%

\begin{abstract}
 We study fairness in collaborative-filtering recommender systems, which are sensitive to discrimination that exists in historical data. Biased data can lead collaborative filtering methods to make unfair predictions against minority groups of users. We identify the insufficiency of existing fairness metrics and propose four new metrics that address different forms of unfairness. These fairness metrics can be optimized by adding fairness terms to the learning objective. Experiments on synthetic and real data show that our new metrics can better measure fairness than the baseline, and that the fairness objectives effectively help reduce unfairness.
\end{abstract}

%


\maketitle


\section{Introduction}

This paper introduces new measures of unfairness in
algorithmic recommendation and demonstrates how
to optimize these metrics to reduce different forms of
unfairness. Since recommender systems make predictions
based on observed data, they can easily inherit
bias that may already exist. To address this issue, we
first describe a process that leads to unfairness in recommender
systems and identify the insufficiency of
demographic parity for this setting. We then propose
four new unfairness metrics that address different forms
of unfairness.
To improve model fairness, we provide five fairness objectives that can be optimized as regularizers. 

We focus on a frequently practiced approach for recommendation called collaborative filtering.
With this approach, predictions are made based on co-occurrence statistics, and most methods assume that the missing ratings are missing at random. Unfortunately, research has shown that sampled ratings have markedly different properties from the users' true preferences \citep{marlin2012collaborative,marlin:recsys09}, 
which is a potential source of unfairness.

We consider a running example of unfair recommendation in education \citep{sacin2009recommendation, thai2010recommender, dascalu2016educational}, in which
the underrepresentation of women in science, technology, engineering, and mathematics (STEM) topics \citep{beede2011women,smith2011women,griffith2010persistence} causes 
the learned model to underestimate women's preferences and be biased towards men. We find this setting a serious motivation to advance understanding of unfairness---and methods to reduce unfairness---in recommendation.

\paragraph{Related Work}
Various studies have considered algorithmic fairness in problems such as classification \citep{pedreshi2008discrimination,lum2016statistical,zafar2017fairness}. 
Removing sensitive features (e.g., gender, race, or age) is often insufficient for fairness. Features are often correlated, so other unprotected attributes can be related to the sensitive features
 \citep{kamishima2011fairness, zemel2013learning}. Moreover, in problems such as collaborative filtering, algorithms do not directly consider measured features and instead infer latent user attributes from their behavior.

Another frequently practiced strategy for encouraging fairness is to enforce \emph{demographic parity}, which is
to ensure that the overall proportion of members in the protected group receiving positive (negative) classifications are identical to the proportion of the population as a whole \citep{zemel2013learning}. 
Based on this non-parity unfairness concept, Kamishima et al.~\citep{kamishima2011fairness, kamishima2012enhancement, kamishima2013efficiency} try to solve the unfairness issue in recommender systems by adding a regularization term that enforces demographic parity. 
However, demographic parity is only appropriate when preferences are unrelated to the sensitive features. In recommendation, user preferences are indeed influenced by sensitive features such as gender, race and age \citep{chausson2010watches, daymont1984job}. 

To address the issues of demographic parity, Hardt et al.~\citep{hardt2016equality} measure unfairness with the true positive rate and true negative rate. They propose that, in a binary setting, given a decision $\hat{Y} \in \{0, 1\}$, a protected attribute $A \in \{0, 1\}$ and the true label  $Y \in \{0, 1\}$, the constraints are \citep{hardt2016equality}
$\Pr\{\hat{Y} =1 | A = 0, Y = y\} = \Pr\{\hat{Y} =1 | A = 1, Y = y\}, y  \in  \{0, 1\}$. 
This idea encourages \emph{equal opportunity} and no longer relies on the assumption of demographic parity, that the target variable is independent of sensitive features. 
Similarly, Calders et al. \citep{calders2013controlling} propose to
impose constrains on the residuals of linear regression
models, which requires not only the mean prediction
but also the mean residuals to be the same across groups.
These ideas form the basis of the unfairness metrics we
propose for recommendation.


\section{Fairness Objectives for Collaborative Filtering}
\label{sec:approach}

This section introduces fairness objectives for collaborative filtering. We begin by reviewing the matrix factorization method. We then describe the various fairness objectives we consider, providing formal definitions and discussion of their motivations.

\subsection{Matrix Factorization}

We consider the task of collaborative filtering using matrix factorization \citep{koren2009matrix}. We have a set of users indexed from 1 to $\numusers$ and a set of items indexed from 1 to $\numitems$. For the $i$th user, let $\group_i$ be a variable indicating which group the $i$th user belongs to. 
For the $j$th item, let $\itemgroup_j$ indicate the item group that it belongs to. 
Let $\rating_{ij}$ be the preference score of the $i$th user for the $j$th item. 

The matrix-factorization formulation assumes that each rating can be represented as
$\rating_{ij} \approx \uservec_i ^\top \itemvec_j + \userbias_i + \itembias_j$,
where $\uservec_i$ is a $d$-dimensional vector representing the $i$th user, $\itemvec_j$ is a $d$-dimensional vector representing the $j$th item, and $\userbias_i$ and $\itembias_j$ are scalar bias terms for the user and item, respectively. The matrix-factorization learning algorithm seeks to learn these parameters from observed ratings $\trainingdata$, typically by minimizing a regularized, squared reconstruction error:
\begin{equation}
\small
J(\usermat, \itemmat, \userbiasvec, \itembiasvec) = \frac{\lambda}{2} \left( ||\usermat||^2_{\mathrm{F}} + ||\itemmat||^2_{\mathrm{F}} \right) + \frac{1}{|\trainingdata|} \sum_{(i, j) \in \trainingdata} \left( \prediction_{ij} - \rating_{ij} \right)^2 ~,
\label{eq:MF-objective}
\end{equation}
where $\userbiasvec$ and $\itembiasvec$ are the vectors of bias terms, and $|| \cdot ||_{\mathrm{F}}$ represents the Frobenius norm. 

\subsection{Fairness Metrics}

We consider a binary group feature distinguishing disadvantaged and advantaged groups. In the STEM example, the disadvantaged group may be women and non-binary gender identities, and the advantaged group may be men.

The first metric is \emph{value unfairness}, which measures inconsistency in signed estimation error across the user types, computed as
\begin{equation}
\small
\metric_\val = \frac{1}{\numitems} \sum_{j = 1}^\numitems \left| \left( \avgpredF_j - \avgrateF_j \right) - \left( \avgpredM_j - \avgrateM_j \right) \right| ~ ,
\label{eq:value-unfairness}
\end{equation}
where $\avgpredF_j$ is the average predicted score for the $j$th item from disadvantaged users, $\avgpredM_j$ is the average predicted score for advantaged users, and $\avgrateF_j$ and $\avgrateM_j$ are the average ratings for the disadvantaged and advantaged users, respectively. 
Value unfairness occurs when one class of user is consistently given higher or lower predictions than their true preferences. 

The second metric is \emph{absolute unfairness}, which measures inconsistency in absolute estimation error across user types, computed as
\begin{equation}
\small
\metric_\absolute = \frac{1}{\numitems} \sum_{j = 1}^\numitems \left| \left| \avgpredF_j - \avgrateF_j \right| - \left| \avgpredM_j - \avgrateM_j \right| \right| ~.
\label{eq:abs-unfairness}
\end{equation}
Absolute unfairness is unsigned, so it captures the quality of prediction for each user type. 

The third metric is \emph{underestimation unfairness}, which measures inconsistency in how much the predictions underestimate the true ratings:
\begin{equation}
\small
\metric_\underest = \frac{1}{\numitems} \sum_{j = 1}^\numitems \left|  \hinge \left(\avgrateF_j - \avgpredF_j \right)- \hinge \left( \avgrateM_j - \avgpredM_j \right) \right| ~,
\label{eq:under-unfairness}
\end{equation}
where $\hinge(x)$ is the hinge function, i.e., $x$ if $x \ge 0$ and 0 otherwise.
Underestimation unfairness is important in settings where missing recommendations are more critical than extra recommendations. 

Conversely, the fourth new metric is \emph{overestimation unfairness}, which measures inconsistency in how much the predictions overestimate the true ratings:
\begin{equation}
\small
\metric_\overest = \frac{1}{\numitems} \sum_{j = 1}^\numitems \left| \hinge\left( \avgpredF_j - \avgrateF_j \right) -  \hinge \left( \avgpredM_j - \avgrateM_j \right) \right| ~.
\label{eq:over-unfairness}
\end{equation}

Finally, a \emph{non-parity} unfairness measure based on the regularization term introduced by Kamishima et al.~\citep{kamishima2011fairness} can be computed as the absolute difference between the overall average ratings of disadvantaged users and that of advantaged users
$\metric_\parity = \left| \avgpredF - \avgpredM \right|$.

To optimize the metric(s), we solve for a local minimum of 
$\min_{\usermat, \itemmat, \userbiasvec, \itembiasvec}
 ~ J(\usermat, \itemmat, \userbiasvec, \itembiasvec) + \alpha
\metric$.

\section{Experiments}
\label{sec:experiments}

We run experiments on simulated course-recommendation data and real movie rating data \citep{harper2016movielens}. 

\subsection{Synthetic data}
\label{sec:synthetic}

In our synthetic experiments, 
we consider four user groups $\group \in \{\textrm{W}, \textrm{WS}, \textrm{M}, \textrm{MS}\}$ and three item groups $\itemgroup \in \{ \textrm{Fem}, \textrm{STEM}, \textrm{Masc} \}$. The user groups represent women who do not enjoy STEM topics (W), women who do enjoy STEM topics (WS), men who do not enjoy STEM topics (M), and men who do (MS). The item groups represent courses that tend to appeal to women (Fem), STEM courses, and courses that tend to appeal to men (Masc). We generate simulated course-recommendation data with two stochastic block models \citep{holland1976local}. Our rating block model determines the probability that a user in a user group likes an item in an item group
\begin{equation}
\tiny
\rateblockmat = \left[
\begin{tabular}{c|ccc}
& Fem & STEM & Masc\\
\midrule
W & 0.8 & 0.2 & 0.2 \\
WS & 0.8 & 0.8 & 0.2 \\
MS & 0.2 & 0.8 & 0.8 \\
M & 0.2 & 0.2 & 0.8
\end{tabular}
\right].
\end{equation}

We use two observation block models that determine the probability a user in a user group rates an item in an item group: one with uniform observation probability for all groups $\takeblockmat^{\textrm{uni}} = [0.4]^{4 \times 3}$ and one with unbalanced observation probabilities inspired by real-world biases
\begin{equation}
\tiny
\takeblockmat^{\textrm{bias}} = \left[
\begin{tabular}{c|ccc}
& Fem & STEM & Masc\\
\midrule
W & 0.6 & 0.2 & 0.1 \\
WS & 0.3 & 0.4 & 0.2 \\
MS & 0.05 & 0.5 & 0.35 \\
M & 0.1 & 0.3 & 0.5 
\end{tabular}
\right]~.
\end{equation}

We define two different user group distributions: one in which each of the four groups is exactly a quarter of the population, and an imbalanced setting where 0.4 of the population is in W, 0.1 in WS, 0.4 in MS, and 0.1 in M. This heavy imbalance is inspired by some of the severe gender imbalance in certain STEM areas today.


\paragraph{Unfairness from different types of underrepresentation}

\begin{figure}[tbp]
\centering
\includegraphics[width=0.23\textwidth]{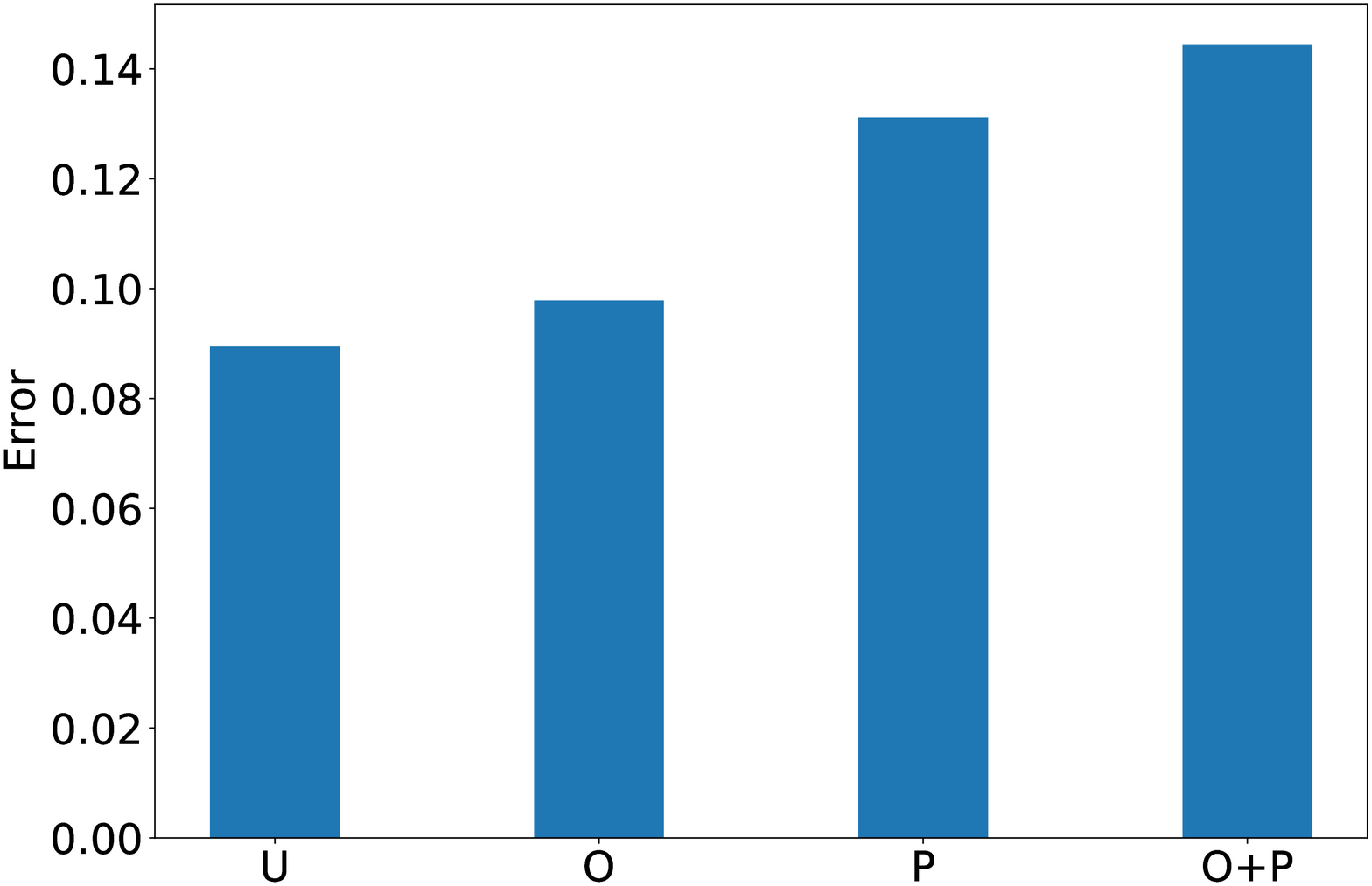}
\includegraphics[width=0.23\textwidth]{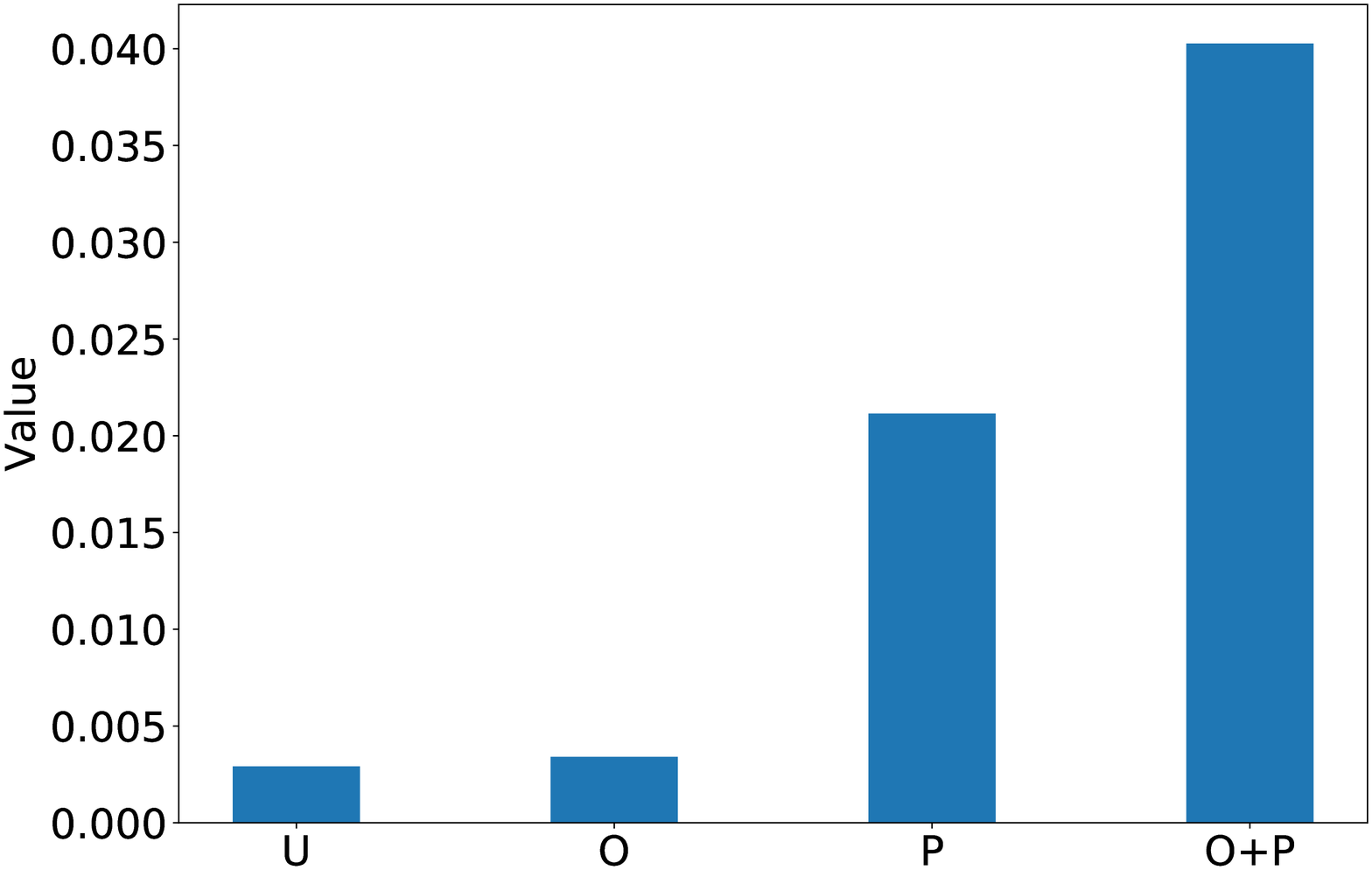}
\includegraphics[width=0.23\textwidth]{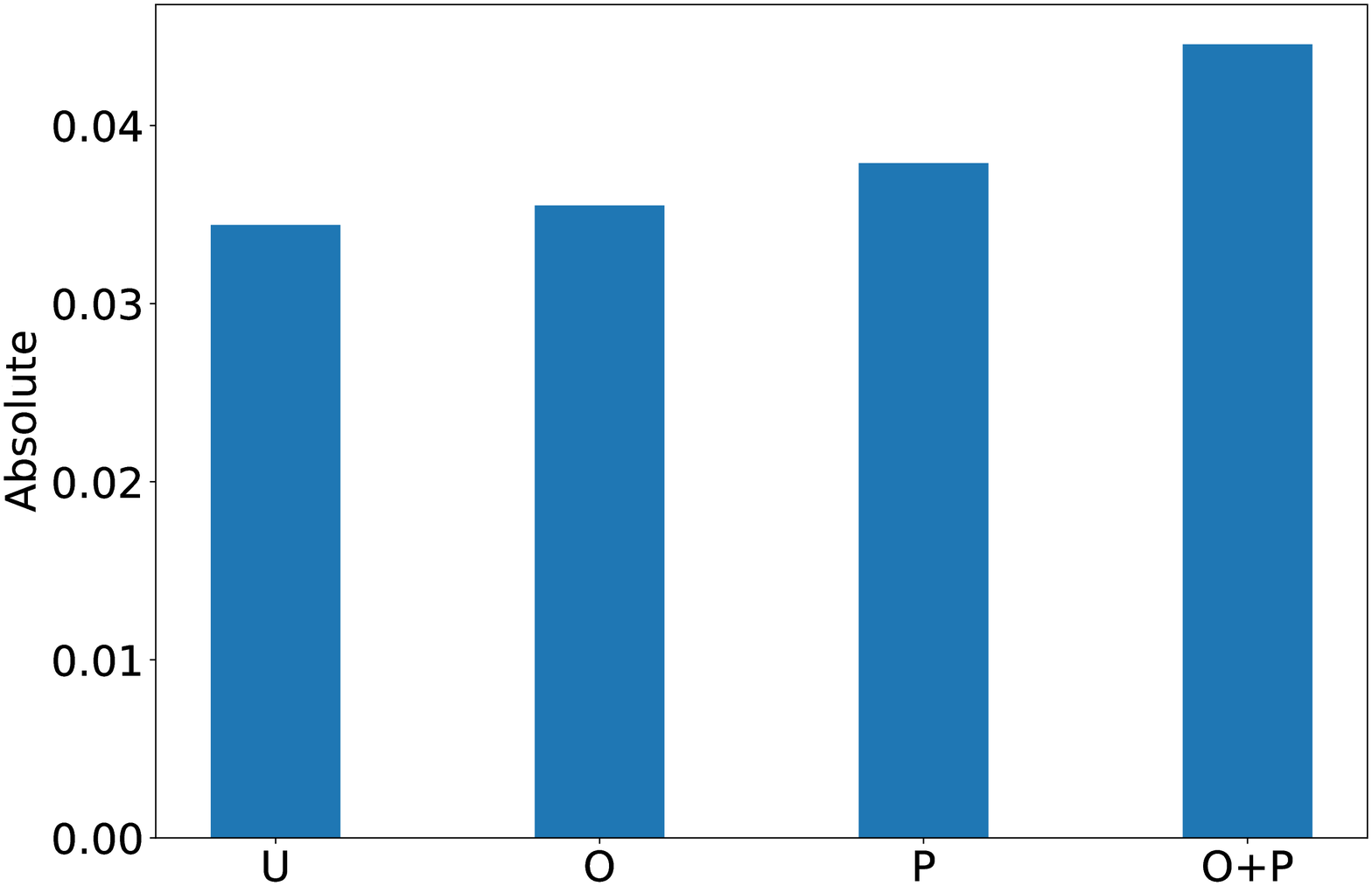}
\includegraphics[width=0.23\textwidth]{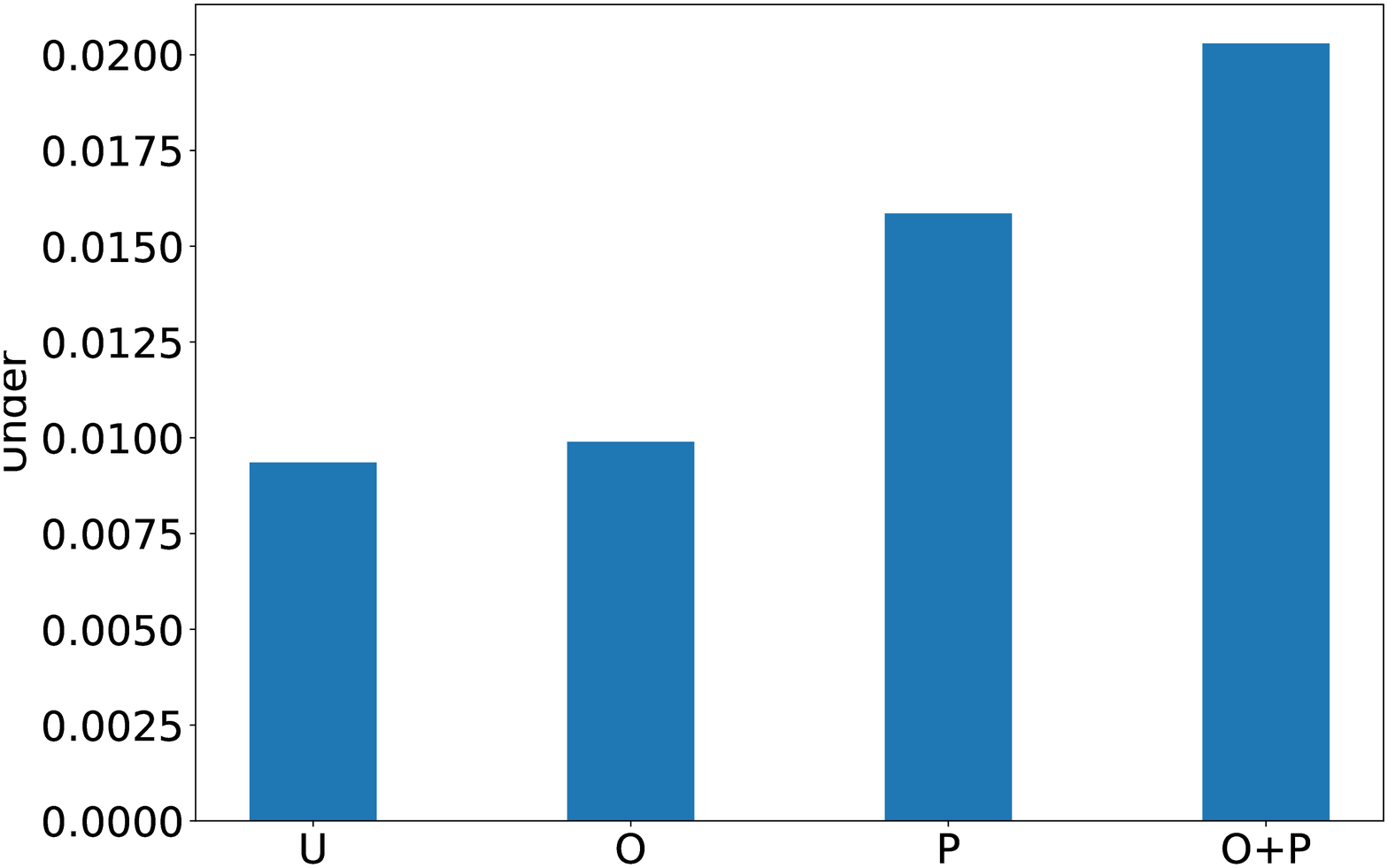}
\includegraphics[width=0.23\textwidth]{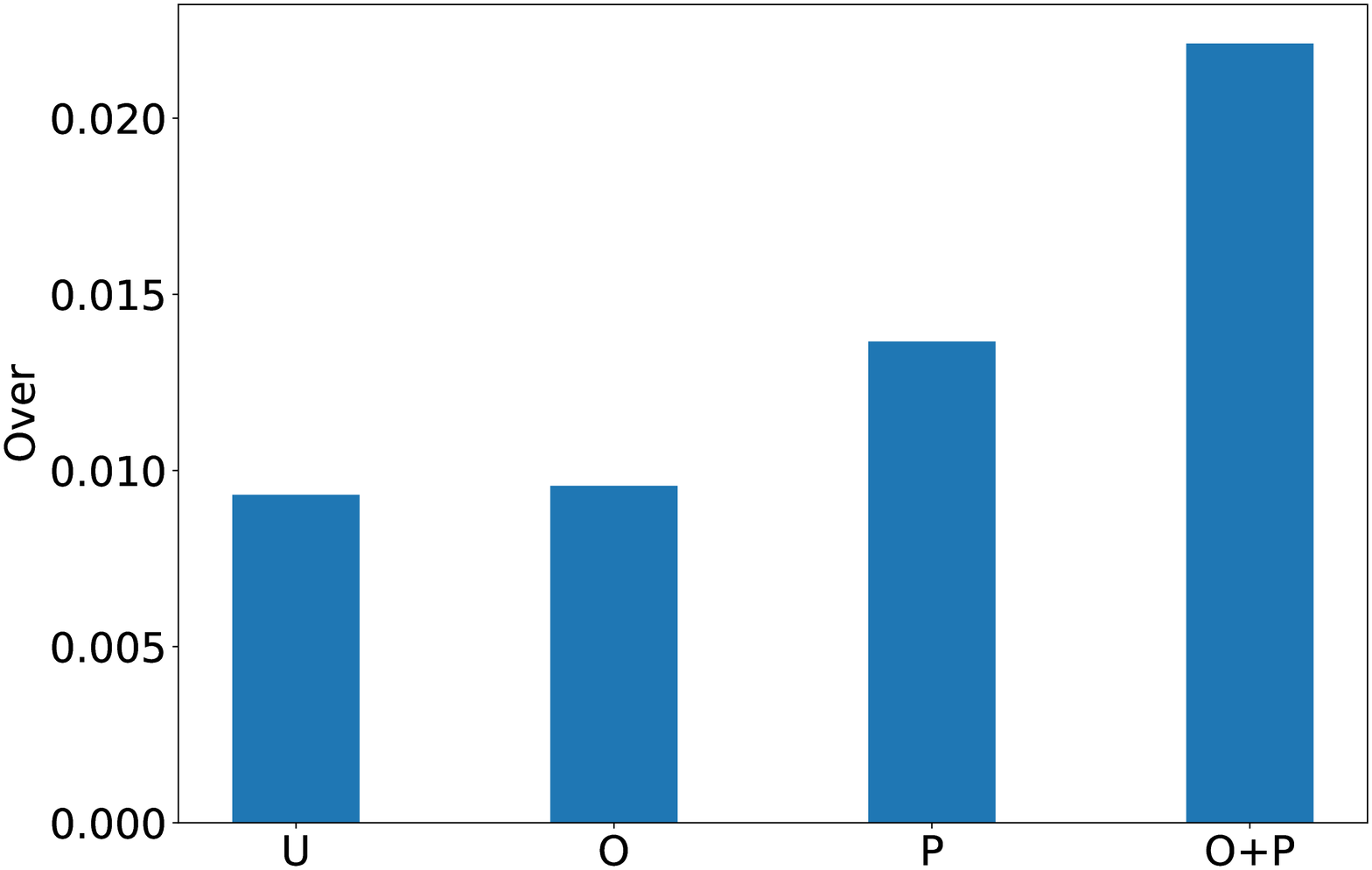}
\includegraphics[width=0.23\textwidth]{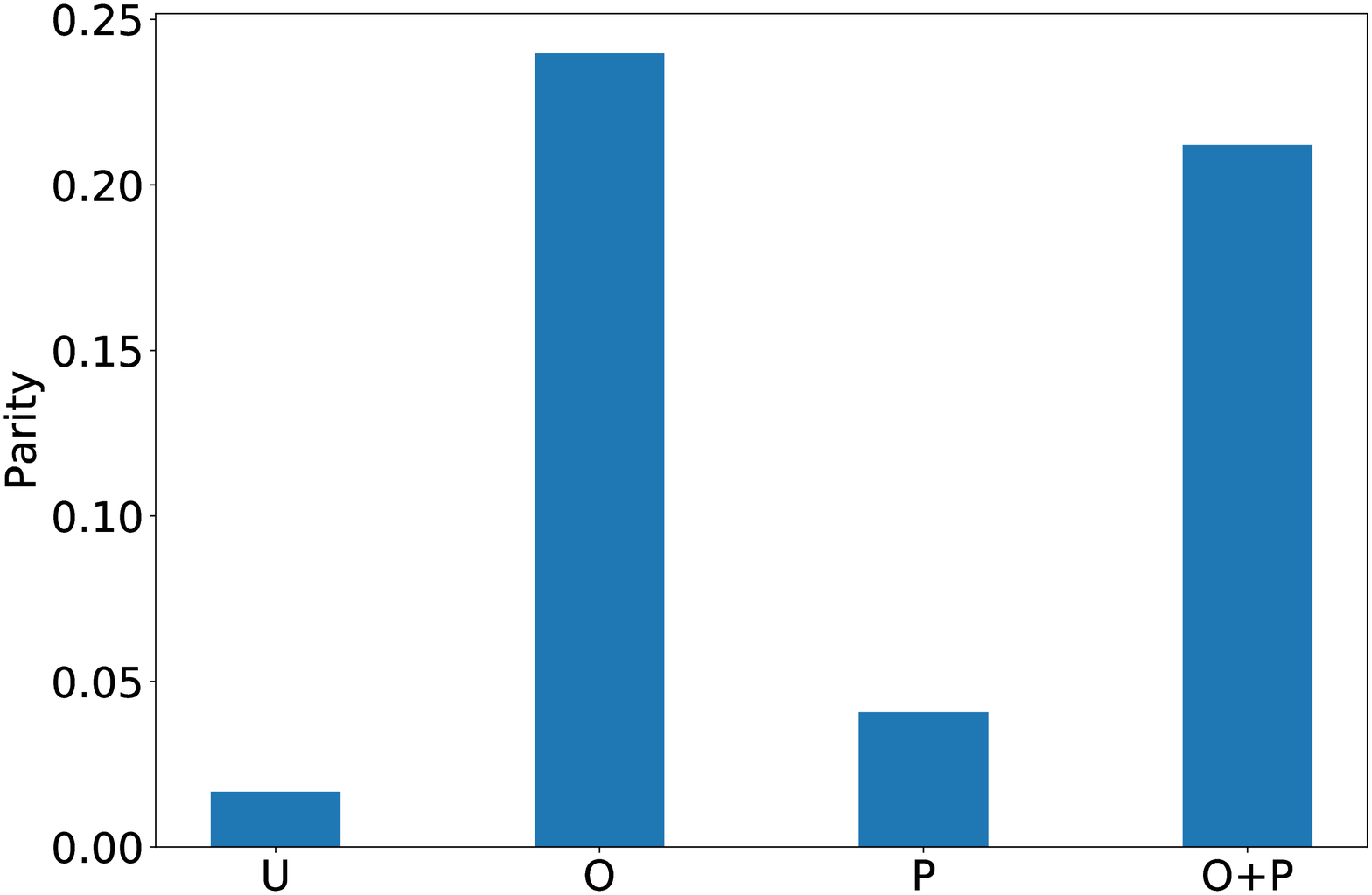}
\caption{Average unfairness scores for standard matrix factorization on synthetic data generated from different underrepresentation schemes.} 
\label{fig:bars}
\end{figure}

Using standard matrix factorization, we measure the various unfairness metrics under the different sampling settings. We average over five random trials and plot the average score in \cref{fig:bars}. In each trial, we generated ratings by 400 users and 300 items with the block models. We label the settings as follows: uniform user groups and uniform observation probabilities (U), uniform groups and biased observation probabilities (O), biased user group populations and uniform observations (P), and biased populations and observations (P+O).

\begin{table*}[tbp]
\caption{Average error and unfairness metrics for synthetic data using different fairness objectives. The best scores and those that are statistically indistinguishable from the best are printed in bold. Each row represents a different unfairness penalty, and each column is the measured metric on the expected value of unseen ratings.}
\label{tab:synthetic}
\centering
{\scriptsize
\begin{tabular}{lllllll}
\toprule
Unfairness & Error & Value & Absolute & Underestimation & Overestimation & Non-Parity \\
\midrule
 None &  0.317 $\pm$ 1.3e-02 & 0.649 $\pm$ 1.8e-02 & 0.443 $\pm$ 2.2e-02 & 0.107 $\pm$ 6.5e-03 & 0.544 $\pm$ 2.0e-02 & 0.362 $\pm$ 1.6e-02 \\
 Value &  \textbf{0.130 $\pm$ 1.0e-02} & \hl{\textbf{0.245 $\pm$ 1.4e-02}} & 0.177 $\pm$ 1.5e-02 & \textbf{0.063 $\pm$ 4.1e-03} & \textbf{0.199 $\pm$ 1.5e-02} & 0.324 $\pm$ 1.2e-02 \\
 Absolute &  0.205 $\pm$ 8.8e-03 & 0.535 $\pm$ 1.6e-02 & \hl{0.267 $\pm$ 1.3e-02} & 0.135 $\pm$ 6.2e-03 & 0.400 $\pm$ 1.4e-02 & 0.294 $\pm$ 1.0e-02 \\
 Under &  0.269 $\pm$ 1.6e-02 & 0.512 $\pm$ 2.3e-02 & 0.401 $\pm$ 2.4e-02 & \hl{\textbf{0.060 $\pm$ 3.5e-03}} & 0.456 $\pm$ 2.3e-02 & 0.357 $\pm$ 1.6e-02 \\
 Over &  \textbf{0.130 $\pm$ 6.5e-03} & 0.296 $\pm$ 1.2e-02 & \textbf{0.172 $\pm$ 1.3e-02} & 0.074 $\pm$ 6.0e-03 & \hl{0.228 $\pm$ 1.1e-02} & 0.321 $\pm$ 1.2e-02 \\
 Non-Parity &  0.324 $\pm$ 1.3e-02 & 0.697 $\pm$ 1.8e-02 & 0.453 $\pm$ 2.2e-02 & 0.124 $\pm$ 6.9e-03 & 0.573 $\pm$ 1.9e-02 & \hl{\textbf{0.251 $\pm$ 1.0e-02}} \\
\bottomrule
\end{tabular}
}
\end{table*}

\begin{table*}[tbp]
\caption{Average error and unfairness metrics for movie-rating data using different fairness objectives.}
\label{tab:movielens}
\centering
{\scriptsize
\begin{tabular}{lllllll}
\toprule
Unfairness & Error & Value & Absolute & Underestimation & Overestimation & Non-Parity \\
\midrule
 None &  0.887 $\pm$ 1.9e-03 & 0.234 $\pm$ 6.3e-03 & 0.126 $\pm$ 1.7e-03 & 0.107 $\pm$ 1.6e-03 & 0.153 $\pm$ 3.9e-03 & 0.036 $\pm$ 1.3e-03 \\
 Value &  0.886 $\pm$ 2.2e-03 & \hl{\textbf{0.223 $\pm$ 6.9e-03}} & 0.128 $\pm$ 2.2e-03 & \textbf{0.102 $\pm$ 1.9e-03} & \textbf{0.148 $\pm$ 4.9e-03} & 0.041 $\pm$ 1.6e-03 \\
 Absolute &  0.887 $\pm$ 2.0e-03 & 0.235 $\pm$ 6.2e-03 & \hl{\textbf{0.124 $\pm$ 1.7e-03}} & 0.110 $\pm$ 1.8e-03 & 0.151 $\pm$ 4.2e-03 & 0.023 $\pm$ 2.7e-03 \\
 Under &  0.888 $\pm$ 2.2e-03 & 0.233 $\pm$ 6.8e-03 & 0.128 $\pm$ 1.8e-03 & \hl{\textbf{0.102 $\pm$ 1.7e-03}} & 0.156 $\pm$ 4.2e-03 & 0.058 $\pm$ 9.3e-04 \\
 Over &  \textbf{0.885 $\pm$ 1.9e-03} & 0.234 $\pm$ 5.8e-03 & \textbf{0.125 $\pm$ 1.6e-03} & 0.112 $\pm$ 1.9e-03 & \hl{\textbf{0.148 $\pm$ 4.1e-03}} & 0.015 $\pm$ 2.0e-03 \\
 Non-Parity &  0.887 $\pm$ 1.9e-03 & 0.236 $\pm$ 6.0e-03 & 0.126 $\pm$ 1.6e-03 & 0.110 $\pm$ 1.7e-03 & 0.152 $\pm$ 3.9e-03 & \hl{\textbf{0.010 $\pm$ 1.5e-03}} \\
\bottomrule
\end{tabular}
}
\end{table*}

The statistics suggest that each underrepresentation type contributes to various forms of unfairness. For all metrics except parity, there is a strict order of unfairness, where uniform data is the most fair
and biasing the populations and observations causes the most unfairness. 
Because of the observation bias, there is actually non-parity in the labeled ratings, so a high non-parity score does not necessarily indicate an unfair situation. 
These tests verify that unfairness can occur with imbalanced populations or observations even when the measured ratings accurately represent user preferences.

\paragraph{Optimization of unfairness metrics}

We optimize fairness objectives under the most imbalanced setting: the user populations are imbalanced, and the sampling rate is skewed. We optimize for 500 iterations of Adam \citep{kingma2014adam}. 


The results are listed in \cref{tab:synthetic}. 
The learning algorithm successfully minimizes the unfairness penalties, generalizing to unseen, held-out user-item pairs. And reducing any unfairness metric does not lead to a significant increase in reconstruction error. The combined objective ``Over+Under'' leads to scores that are close to the minimum of each metric except parity. 



\subsection{Real data}

We use the Movielens Million Dataset \citep{harper2016movielens}, which contains ratings in [1,5] by 6,040 users and 3,883 movies. 
We manually selected five genres (\emph{action}, \emph{crime}, \emph{musical}, \emph{romance}, and \emph{sci-fi}) that each have different forms of gender imbalance and only consider movies that list these genres. Then we filtered the users to only consider those who rated at least 50 of the selected movies.
After filtering by genre and rating frequency, we have 2,953 users and 1,006 movies in the dataset.


We run three trials in which we randomly split the ratings into training and testing sets,
the average scores are listed in \cref{tab:movielens}.
As in the synthetic setting, the results show that optimizing each unfairness metric leads to the best performance on that metric without a significant change in the reconstruction error. 

\section{Conclusion}
\label{sec:conclusion}

In this paper, we discussed various types of unfairness that can occur in collaborative filtering. We demonstrate that these forms of unfairness can occur even when the observed rating data accurately reflects the users' preferences. We propose four fairness metrics and demonstrate that augmenting matrix factorization objectives with these metrics as penalty functions enables their minimization. Our experiments on synthetic and real data show that minimization of these unfairness metrics is possible with no significant increase in reconstruction error.
However, no single objective was the best for all unfairness metrics, so it remains necessary for practitioners to consider precisely which form of unfairness is most important in their application and optimize that specific objective.

\paragraph{Future Work}

While our work here focused on improving fairness among user groups, we did not address fair treatment of different item groups. The model could be biased towards certain items, e.g., performing better for some items than others. Achieving fairness for both user and items may be important when considering that the items may also suffer from discrimination or bias, e.g., when courses are taught by instructors with different demographics. 

Moreover, our fairness metrics assume that users rate items according to their true preferences. This assumption is likely violated in real data, since ratings can also be influenced by environmental factors. E.g., in education, a student's rating for a course also depends on whether the course has an inclusive and welcoming learning environment. However, addressing this type of bias may require additional information or external interventions beyond the provided rating data.

Finally, we are investigating methods to reduce unfairness by directly modeling the two-stage sampling process we used in \cref{sec:synthetic}. Explicitly modeling the rating and observation probabilities as separate variables may enable a principled, probabilistic approach to address these forms of data imbalance.

\bibliographystyle{plain}
\bibliography{yao-nips17} 

\begin{thebibliography}{10}

\bibitem{beede2011women}
David~N Beede, Tiffany~A Julian, David Langdon, George McKittrick, Beethika
  Khan, and Mark~E Doms.
\newblock Women in {STEM}: A gender gap to innovation.
\newblock {\em U.S. Department of Commerce, Economics and Statistics
  Administration}, 2011.

\bibitem{calders2013controlling}
Toon Calders, Asim Karim, Faisal Kamiran, Wasif Ali, and Xiangliang Zhang.
\newblock Controlling attribute effect in linear regression.
\newblock In {\em Data Mining (ICDM), 2013 IEEE 13th International Conference
  on}, pages 71--80. IEEE, 2013.

\bibitem{chausson2010watches}
Olivia Chausson.
\newblock Who watches what? {A}ssessing the impact of gender and personality on
  film preferences.
\newblock {\em
  http://mypersonality.org/wiki/doku.php?id=movie\_tastes\_and\_personality},
  2010.

\bibitem{dascalu2016educational}
Maria-Iuliana Dascalu, Constanta-Nicoleta Bodea, Monica~Nastasia Mihailescu,
  Elena~Alice Tanase, and Patricia Ordo{\~n}ez~de Pablos.
\newblock Educational recommender systems and their application in lifelong
  learning.
\newblock {\em Behaviour \& Information Technology}, 35(4):290--297, 2016.

\bibitem{daymont1984job}
Thomas~N. Daymont and Paul~J. Andrisani.
\newblock Job preferences, college major, and the gender gap in earnings.
\newblock {\em Journal of Human Resources}, pages 408--428, 1984.

\bibitem{griffith2010persistence}
Amanda~L. Griffith.
\newblock Persistence of women and minorities in {STEM} field majors: Is it the
  school that matters?
\newblock {\em Economics of Education Review}, 29(6):911--922, 2010.

\bibitem{hardt2016equality}
Moritz Hardt, Eric Price, Nati Srebro, et~al.
\newblock Equality of opportunity in supervised learning.
\newblock In {\em Advances in Neural Information Processing Systems}, pages
  3315--3323, 2016.

\bibitem{harper2016movielens}
F~Maxwell Harper and Joseph~A Konstan.
\newblock The {M}ovielens datasets: History and context.
\newblock {\em ACM Transactions on Interactive Intelligent Systems (TiiS)},
  5(4):19, 2016.

\bibitem{holland1976local}
Paul~W. Holland and Samuel Leinhardt.
\newblock Local structure in social networks.
\newblock {\em Sociological Methodology}, 7:1--45, 1976.

\bibitem{kamishima2012enhancement}
Toshihiro Kamishima, Shotaro Akaho, Hideki Asoh, and Jun Sakuma.
\newblock Enhancement of the neutrality in recommendation.
\newblock In {\em Proceedings of the 2nd Workshop on Human Decision Making in
  Recommender Systems (Decisions@RecSys)}, pages 8--14, 2012.

\bibitem{kamishima2013efficiency}
Toshihiro Kamishima, Shotaro Akaho, Hideki Asoh, and Jun Sakuma.
\newblock Efficiency improvement of neutrality-enhanced recommendation.
\newblock In {\em Proceedings of the 3rd Workshop on Human Decision Making in
  Recommender Systems (Decisions@RecSys)}, pages 1--8, 2013.

\bibitem{kamishima2011fairness}
Toshihiro Kamishima, Shotaro Akaho, and Jun Sakuma.
\newblock Fairness-aware learning through regularization approach.
\newblock In {\em 11th International Conference on Data Mining Workshops
  (ICDMW)}, pages 643--650. IEEE, 2011.

\bibitem{kingma2014adam}
Diederik Kingma and Jimmy Ba.
\newblock Adam: A method for stochastic optimization.
\newblock {\em arXiv preprint arXiv:1412.6980}, 2014.

\bibitem{koren2009matrix}
Yehuda Koren, Robert Bell, and Chris Volinsky.
\newblock Matrix factorization techniques for recommender systems.
\newblock {\em Computer}, 42(8), 2009.

\bibitem{lum2016statistical}
Kristian Lum and James Johndrow.
\newblock A statistical framework for fair predictive algorithms.
\newblock {\em arXiv preprint arXiv:1610.08077}, 2016.

\bibitem{marlin2012collaborative}
Benjamin Marlin, Richard~S Zemel, Sam Roweis, and Malcolm Slaney.
\newblock Collaborative filtering and the missing at random assumption.
\newblock {\em arXiv preprint arXiv:1206.5267}, 2012.

\bibitem{marlin:recsys09}
Benjamin~M. Marlin and Richard~S. Zemel.
\newblock Collaborative prediction and ranking with non-random missing data.
\newblock In {\em Proceedings of the Third ACM Conference on Recommender
  Systems}, pages 5--12. ACM, 2009.

\bibitem{pedreshi2008discrimination}
Dino Pedreshi, Salvatore Ruggieri, and Franco Turini.
\newblock Discrimination-aware data mining.
\newblock In {\em Proceedings of the 14th ACM SIGKDD International Conference
  on Knowledge Discovery and Data Mining}, pages 560--568. ACM, 2008.

\bibitem{sacin2009recommendation}
Cesar~Vialardi Sacin, Javier~Bravo Agapito, Leila Shafti, and Alvaro Ortigosa.
\newblock Recommendation in higher education using data mining techniques.
\newblock In {\em Educational Data Mining}, 2009.

\bibitem{smith2011women}
Emma Smith.
\newblock Women into science and engineering? {G}endered participation in
  higher education {STEM} subjects.
\newblock {\em British Educational Research Journal}, 37(6):993--1014, 2011.

\bibitem{thai2010recommender}
Nguyen Thai-Nghe, Lucas Drumond, Artus Krohn-Grimberghe, and Lars
  Schmidt-Thieme.
\newblock Recommender system for predicting student performance.
\newblock {\em Procedia Computer Science}, 1(2):2811--2819, 2010.

\bibitem{zafar2017fairness}
Muhammad~Bilal Zafar, Isabel Valera, Manuel Gomez~Rodriguez, and Krishna~P.
  Gummadi.
\newblock Fairness constraints: Mechanisms for fair classification.
\newblock {\em arXiv preprint arXiv:1507.05259}, 2017.

\bibitem{zemel2013learning}
Rich Zemel, Yu~Wu, Kevin Swersky, Toni Pitassi, and Cynthia Dwork.
\newblock Learning fair representations.
\newblock In {\em Proceedings of the 30th International Conference on Machine
  Learning}, pages 325--333, 2013.

\end{thebibliography}

\end{document}